\documentclass[acmsmall,screen]{acmart}

\AtBeginDocument{%
  \providecommand\BibTeX{{%
    \normalfont B\kern-0.5em{\scshape i\kern-0.25em b}\kern-0.8em\TeX}}}

\setcopyright{acmcopyright}
\copyrightyear{2018}
\acmYear{2018}
\acmDOI{10.1145/1122445.1122456}

\acmConference[Woodstock '18]{Woodstock '18: ACM Symposium on Neural
  Gaze Detection}{June 03--05, 2018}{Woodstock, NY}
\acmBooktitle{Woodstock '18: ACM Symposium on Neural Gaze Detection,
  June 03--05, 2018, Woodstock, NY}
\acmPrice{15.00}
\acmISBN{978-1-4503-XXXX-X/18/06}

\setcopyright{none}
\renewcommand\footnotetextcopyrightpermission[1]{} 

\usepackage{amsmath,amsfonts}
\usepackage{multirow}
\usepackage{algorithm}
\usepackage{algorithmic}
\usepackage{graphicx}
\usepackage{textcomp}
\usepackage{xcolor}
\usepackage{framed}
\usepackage{listings}
\usepackage{bbding} 
\usepackage{color}
\usepackage{enumitem}
\usepackage{threeparttable}
\usepackage{array}
\usepackage{xspace}

\usepackage{graphicx}    
\usepackage{float} 
\usepackage[most]{tcolorbox}
\usepackage{code_styles/style}
\usepackage{wrapfig}
\usepackage{makecell}

\newcounter{finding}
\newcommand{\finding}[1]{
  \begin{tcolorbox}[enhanced, left=3mm,right=3mm,
    colback=gray!10, colframe=gray!80, boxrule=0pt,
    borderline west={4pt}{0pt}{gray!90},
    breakable
    ]
    \textbf{Finding \Roman{finding}:} #1
    \stepcounter{finding}
    \end{tcolorbox}
}

\usepackage{colortbl}

\usepackage{booktabs}

\newcommand{\sssec}[1]{\vspace*{0.02in}\noindent\textbf{#1}}
\newcolumntype{M}[1]{>{\centering\arraybackslash}p{#1}}

\usepackage{tikz}
\DeclareRobustCommand*\circled[1]{\tikz[baseline=(char.base)]{ \node[shape=circle,draw,color=white,fill=black,inner sep=0.5pt] (char){#1};}}

\begin{document}

\author{Peiran Wang}
\authornote{Both authors contributed equally to this research.}
\affiliation{%
  \institution{UCLA}
  \city{Los Angeles}
  \state{CA}
  \country{USA}
}
\email{peiranwang@ucla.edu}

\author{Ying Li}
\authornotemark[1]
\affiliation{%
  \institution{UCLA}
  \city{Los Angeles}
  \state{CA}
  \country{USA}
}
\email{yinglee@ucla.edu}

\author{Yuqiang Sun}
\affiliation{%
  \institution{NTU}
  \city{Singapore}
  \country{Singapore}
}
\email{yuqiang.sun@ntu.edu.sg}

\author{Chengwei Liu}
\affiliation{%
  \institution{NTU}
  \city{Singapore}
  \country{Singapore}
}
\email{chengwei.liu@ntu.edu.sg}

\author{Yang Liu}
\affiliation{%
  \institution{NTU}
  \city{Singapore}
  \country{Singapore}
}
\email{yangliu@ntu.edu.sg}

\author{Yuan Tian}
\affiliation{%
  \institution{UCLA}
  \city{Los, Angeles}
  \country{USA}
}
\email{yuant@ucla.edu}

\title{From Docs to Descriptions: Smell-Aware Evaluation of MCP Server Descriptions}

\renewcommand{\shortauthors}{Zhang, et al.}

\begin{abstract}

The Model Context Protocol (MCP) has rapidly become a de facto standard for connecting LLM-based agents with external tools via reusable MCP servers. In practice, however, server selection and onboarding rely heavily on free-text tool descriptions that are intentionally loosely constrained. Although this flexibility largely ensures the scalability of MCP servers, it also creates a reliability gap that descriptions often misrepresent or omit key semantics, increasing trial-and-error integration, degrading agent behavior, and potentially introducing security risks.
To this end, we present the first systematic study of \textbf{description smells} in MCP tool descriptions and their impact on usability. Specifically, we synthesize software/API documentation practices and agentic tool-use requirements into a four-dimensional quality standard—accuracy, functionality, information completeness, and conciseness, covering 18 specific smell categories. Using this standard, we conducted a large-scale empirical study on a well-constructed dataset of 10,831 MCP servers. We find that description smells are pervasive (e.g., 73\% repeated tool names; thousands with incorrect parameter semantics or missing return descriptions), reflecting a “code-first, description-last” pattern. Through a controlled mutation-based study, we show these smells significantly affect LLM tool selection, with functionality and accuracy having the largest effects (+11.6\% and +8.8\%, p < 0.001). In competitive settings with functionally equivalent servers, standard-compliant descriptions reach 72\% selection probability (260\% over a 20\% baseline), demonstrating that smell-guided remediation yields substantial practical benefits. We release our labeled dataset and standards to support future work on reliable and secure MCP ecosystems.

\end{abstract}




\maketitle

\section{Introduction}\label{sec:intro}

The Model Context Protocol (MCP)~\cite{mcp}, proposed in 2025, has rapidly become a de facto agreement to standardize how LLM-based agents (clients) connect to external tools and data providers (servers), allowing tool capabilities to be shared and reused across ecosystems rather than implemented as one-off integrations. Since its public introduction, MCP has gained strong momentum in both research and industry communities, with a fast-growing ecosystem of MCP servers, registries, and integration toolchains that lower the cost of tool onboarding and accelerate agent deployment at scale~\cite{mcp-official}. For instance, large enterprises and platforms, such as Google, Microsoft, GitHub, etc., have also quickly reacted to boost the integration of their products with AI Agents. 

However, after the prosperity of the MCP ecosystem, the main concerns of the community are gradually shifting from \textit{can we integrate} to \textit{can we reliably choose and bring onboard the right server}. 
Specifically, MCP servers expose \textit{tools}, \textit{resources}, and \textit{prompts}, and clients (such as LLM-based agents) can discover, invoke, and integrate into their reasoning workflows, by the free-textual MCP descriptions~\cite{mcp-official}. 
To facilitate reuse across diverse ecosystems, MCP keeps the description layer loosely constrained, so that users can be provided with maximized flexibility to design and incorporate MCP servers as they wish. 
However, this flexibility is also a double-edged sword, the less-constrained MCP descriptions also brings new problem that, as the main basis for selection and on-boarding MCP servers, MCP server providers are allowed to arbitrarily write the descriptions of tools, thus, mismatches between what users or agents expect and what a server actually supports become common, which could largely increase the cost for trial-and-error integration and hinder the effective reuse of MCP servers, and even introduce security risks, such as tool abuse, privilege escalation, and data exfiltration.

To this end, some existing works have pioneered their research in security and performance in the MCP ecosystems. For instance, Wang et al.~\cite{mpma-attack} have identified manipulation attacks through malicious description editing, Radosevich et al.~\cite{mcp-safety} have audited the unsafe MCP servers, and Wang et al.~\cite{mcp-bench} have developed benchmark suites to evaluate agent performance across diverse MCP servers. Some researchers have also investigated the security threats hidden in other LLM-interactive texts (i.e., prompts), such as prompt injection attacks~\cite{wu2024isolategpt, zhan2024injecagent, iqbal2024llm, fu2024imprompter, debenedetti2024agentdojo, hines2024defending, bagdasarian2024airgapagent} and generative engine optimization~\cite{aggarwal2024geo, rejon2025generative}.
However, to the best of our knowledge, as the most critical basis of MCP, no existing research has systematically investigated the quality of MCP tool descriptions and to what extent this can influence the performance of user systems.
To this end, we aim to conduct a comprehensive study to demystify the bad smells of MCP descriptions (i.e., similar to code smell), and evaluate their influence to MCP usability, as well as possible mitigation solutions.

To achieve this, we face three major challenges.
\textbf{C1: Lack of Standard.} As an emerging artifact for third-party integration, MCP server descriptions lack well-established standards comparable to those for traditional software engineering documentation. In addition, descriptions are typically short and heterogeneous, which makes it difficult to operationalize what constitutes \textit{good} quality without a principled reference framework.
\textbf{C2: Influence Is Hard to Capture.} The impact of MCP descriptions is primarily behavioral, because their effects are mediated by how human users and LLM agents interpret the text and act on it during server selection, onboarding, and use. This mediation makes it challenging to quantify influence using superficial text-based metrics alone.
\textbf{C3: Actionability of Remediation.} Even when quality issues or smells can be detected, translating them into concrete and reliable improvements is non-trivial. Remediation must remain consistent with actual server behavior, avoid overstating capabilities, and be robust to future version changes, otherwise revisions may introduce new misunderstandings or maintenance burden.

To this end, in this paper, we conduct a comprehensive study to investigate the bad smells in MCP descriptions and their influence to usability. To come up with a systematic evaluation framework for MCP descriptions, considering that there are mature research works and evaluation practices in the field of traditional SE documentation, such as API documentation, for \textbf{C1}, we first conducted a systematic literature review to collect and construct a taxonomy of documentation evaluation properties in traditional SE documentation as the fundamental evaluation model for MCP description. Next, based on the identified properties, we look into current MCP servers published in mainstream MCP markets, to demystify the bad smells that exist in MCP server implementations, then we derive the taxonomy of perspectives that fit the scenario of MCP servers based on them. 
After that, we conduct the empirical study to investigate the perspectives and to what extent do MCP descriptions are poor upon in the context of bad smells. 
Third, considering the huge difference between traditional SE documentation and MCP tool description, the bad smells in traditional views may not fit in the scenario of MCP, for \textbf{C2}, we further conducted a contrast study to investigate how these identified smells can influence the performance of MCP tools, by mutating MCP tool descriptions and measuring the behavioral changes. Moreover, based on these identified influential smells, we further investigate can these smells effectively guide the improvement and bring the competitive advantages of MCP servers in real-world scenarios.

Specifically, we conduct this study by answering the following research questions:
\begin{itemize}[leftmargin=*]
    \item \textbf{RQ1:Taxonomy Analysis.} What are the taxonomy of perspectives of bad smells in the quality of MCP tool descriptions? 

\item \textbf{RQ2: Prevalence Analysis.} How prevalent and what are the distributions of these bad smells in the MCP ecosystems? 

\item \textbf{RQ3: Influence Analysis.} To what extent can these bad smells in the MCP tool descriptions influence the usability?

\item \textbf{RQ4: Mitigation Effectiveness.} To what extent can these bad smells guide the improvement of MCP tool description and enhance competitive advantages for MCP servers?
\end{itemize}

Our study has also revealed some interesting findings. 
First, after going through 7 major types of quality standards that are suitable for MCP servers and their existence in the wild, we identified 4 core dimensions (accuracy, functionality, completeness, and conciseness), including 18 specific categories of description smells in the MCP ecosystem. Next, these description smells are pervasive in the current MCP ecosystem. Among 10,831 studied MCP servers, 73\% suffer from repeated tool names, 3,449 exhibit wrong parameter meanings, and 3,093 lack return value descriptions, reflecting a widespread ``Code-First, Description-Last'' development pattern. Moreover, our experiment confirmed that all four dimensions have various impact for LLM tool selection, with functionality and accuracy being the most influential for tool selection (+11.6\% and +8.8\%, respectively, $p < 0.001$). Furthermore, in real-world competitive scenarios with functionally equivalent servers, a standard-compliant server achieves a 72\% selection probability (260\% increase over the 20\% baseline), demonstrating that description quality provides substantial competitive advantage, which indicate the effectiveness of the standards on improving MCP descriptions.

In summary, we conclude the contribution of this paper as follows:

\noindent $\bullet$ By synthesizing software documentation standards, API documentation, and agentic framework requirements, we derived a four-dimensional quality standard (accuracy, functionality, information completeness, and conciseness) comprising 18 specific smell categories for MCP tool descriptions.

\noindent $\bullet$ We constructed a large-scale MCP server dataset (over 10K MCP servers) that have been well-labeled with these specific smells in their tool descriptions for further research. We have open-sourced it at~\cite{anonymous2026from}.

\noindent $\bullet$ We investigated the prevalence and influence of description smells in real-world MCP servers, and our findings has been proved that they can be effectively utilized to improve the quality of MCP descriptions and bring competitive advantages to MCP servers.
\section{Preliminaries}\label{sec:back}
\subsection{MCP Workflow}\label{sec:back:ecosystem}
\begin{figure}[]
    \centering
    \includegraphics[width=0.8\linewidth]{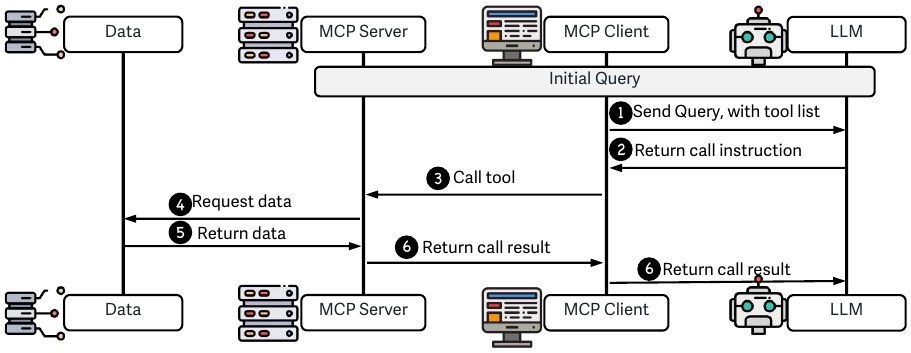}
    \caption{The workflow of the MCP pipeline}
    \label{fig:mcp-pipeline}
    \footnotesize
\end{figure}
The Model Context Protocol (MCP) establishes a unified interaction protocol that connects MCP clients, LLMs, MCP servers, and external data sources within a standardized ecosystem. 
Fig. \ref{fig:mcp-pipeline} illustrates the overall workflow, showing how a client initiates a query that is processed collaboratively by the MCP server and the LLM.
Here, we first introduce the entities involved: 

\sssec{MCP client}.
The client represents the user-facing endpoint (e.g., a conversational agent interface) responsible for initiating the query.
It transmits an initial request along with a list of available tools (including tool names and descriptions) provided by the MCP server. 
The client acts as both the query originator and the result aggregator, managing the lifecycle of the interaction.

\sssec{MCP server}.
The MCP server acts as the central registry and execution hub for tools. 
It exposes tool metadata, including name and description, which is used by the LLM to determine which tool to call. 
Upon receiving a tool call request, the MCP server executes the corresponding tool and returns the execution result. 

\sssec{LLM}.
The LLM functions as the control center of the ecosystem. 
Given the client's query and the MCP server's available tool descriptions, it interprets the user intent and selects the appropriate tool to complete the user intent.
Once it accumulates enough information, the LLM synthesizes the final response for the client.

\sssec{Data}.
The data source represents the external environment (e.g., databases, APIs, or file systems) accessed through the MCP servers' tools. 
Each tool call may trigger one or more data retrieval operations, which are returned to the LLM through the MCP server.

Then, we introduce the whole workflow:
\begin{enumerate}[noitemsep, topsep=1pt, leftmargin=*, label={\circled{\arabic*}}]
    \item The MCP client initiates the process by sending an initial query and a tool list to the LLM. 
    \item The LLM processes the query and returns a tool call instruction to the MCP client. 
    \item The MCP client calls the corresponding tool on the MCP server. 
    \item The MCP server requests data from external data sources. 
    \item The data source returns the requested information to the MCP server. 
    \item The MCP server sends the execution result back to the MCP client, which is then forwarded to the LLM to synthesize the final response.
\end{enumerate}

\subsection{MCP Ecosystem}\label{sec:back:market}


Unlike traditional API or package marketplaces that are tightly controlled by a few providers, the MCP environment is open: anyone can register and publish a server as long as it follows the MCP specification.
Moreover, there are also many similar protocols (e.g., Semantic Kernel by Microsoft \cite{microsoft2024semantic}, function call format by OpenAI \cite{openai2023function}, and LangChain framework \cite{harrison2023langchain}).
Among them, MCP is the most widely used one, attracting thousands of developers.
Each MCP server exposes one or more tools, functions, resources, or prompts that can be discovered and called by LLM-based agents or other clients. 

\begin{figure}[]
    \centering
    \includegraphics[width=0.9\linewidth]{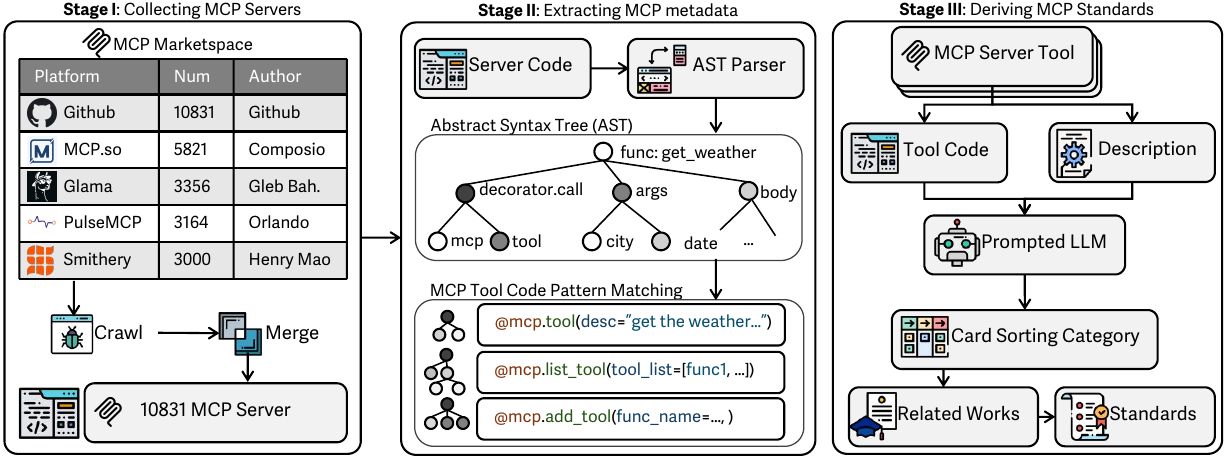}
    \caption{
Overview of the study.
    }
    \label{fig:workflow}
\end{figure}

\sssec{Where to find MCP servers?}
As summarized in Fig. \ref{fig:workflow} (left side), the MCP ecosystem has already surpassed ten thousand public servers across multiple hosting platforms, including GitHub, MCP.so, Glama, PulseMCP, and Smithery, etc.
Among these, GitHub maintains an open repository of all MCP servers, while other platforms like MCP.so and PulseMCP serve as dedicated hubs that propagate the MCP servers to the users and redirect users to the specific repository on GitHub. 

\sssec{Who can upload MCP servers?}
Any developer, research group, or organization can upload and register an MCP server. 
In practice, most servers are open-source implementations contributed by independent developers or research teams, who integrate external APIs (e.g., weather, calendar, database) and expose them via standardized MCP decorators or registration calls (e.g., \texttt{@mcp.tool()}, \texttt{server.addTool()}). 
Enterprises and cloud providers also participate by publishing product-linked MCP servers to these platforms (e.g., Slack, Google Maps, etc.).
They will also maintain their own MCP platforms, and these enterprise-served platforms provide service to the enterprises with their own products, etc.
Such openness encourages rapid expansion of the MCP server's scale, but also introduces heterogeneity in description quality and compliance (e.g., company providers' MCP servers and personal contributors' MCP servers may share vastly different quality and standards), motivating the standardization effort described later in this paper.

\sssec{Who can use MCP servers?}
On the consumer side, MCP servers are primarily accessed by LLM agents, workflow frameworks, and end users through MCP-compatible clients. 
Developers integrate these servers into their own agent system, while non-technical users interact indirectly, via conversational agents or automation systems (e.g., Claude computer-use agent) that rely on these servers for tool calls.



\subsection{Problem Statement}\label{sec:problem}

The MCP architecture, as depicted in Fig.~\ref{fig:mcp-pipeline}, establishes a computational workflow wherein the LLM is tasked with interpreting user queries, selecting appropriate tools from the set exposed by MCP servers, and constructing syntactically and semantically valid invocations predicated upon tool metadata.
Formally, let $\mathcal{T} = \{t_1, t_2, \ldots, t_n\}$ denote the set of tools exposed by MCP servers, where each tool $t_i$ is characterized by a tuple:
\begin{equation}
t_i = \langle \textit{name}_i, \textit{desc}_i, \textit{schema}_i, \textit{impl}_i \rangle
\end{equation}
where $\textit{name}_i \in \Sigma^*$ represents the tool identifier, $\textit{desc}_i \in \Sigma^*$ denotes the natural language description, $\textit{schema}_i$ specifies the input parameter schema, and $\textit{impl}_i$ represents the underlying implementation.
A fundamental architectural constraint is that the LLM operates under conditions of incomplete information: access to $\textit{impl}_i$ is precluded, and decisions must be derived solely from the observable metadata $\mathcal{M}_i = \langle \textit{name}_i, \textit{desc}_i, \textit{schema}_i \rangle$.
This information asymmetry elevates tool descriptions to a position of singular importance within the interaction pipeline.

\noindent\textbf{Tool Selection as a Critical Bottleneck.}
Tool selection constitutes the initial phase of the MCP workflow, the failure of which renders all downstream processing ineffectual.
Given a user query $q \in \mathcal{Q}$ and the observable metadata of available tools, the LLM implements a selection function:
\begin{equation}
f_{\text{select}}: \mathcal{Q} \times 2^{\mathcal{M}} \rightarrow \mathcal{T} \cup \{\bot\}
\end{equation}
where $\bot$ denotes the null selection (no tool invoked).
The selection decision is computed as:
\begin{equation}
t^* = \operatorname*{argmax}_{t_i \in \mathcal{T}} P(t_i \mid q, \mathcal{M}_i; \theta_{\text{LLM}})
\end{equation}
where $\theta_{\text{LLM}}$ parameterizes the language model.
Critically, since $\textit{impl}_i$ is unobservable, the probability $P(t_i \mid q, \mathcal{M}_i)$ is predominantly determined by the semantic alignment between $q$ and $\textit{desc}_i$.
This constraint presents non-trivial challenges across the spectrum of tool selection architectures---whether direct semantic selection, retrieval-augmented generation, or hierarchical indexing---as all approaches exhibit inherent dependence on the semantic fidelity of $\textit{desc}_i$.

\noindent\textbf{Failure Modes Induced by Description Deficiencies.}
Let $t^*$ denote the ground-truth optimal tool for query $q$, and let $\hat{t} = f_{\text{select}}(q, \mathcal{M})$ denote the selected tool.
Inadequate tool descriptions precipitate two principal categories of operational failures:

\noindent\textit{Failure I} (Missed Invocation): occurs when the correct tool exists but is not selected:
\begin{equation}
\mathcal{F}_I: t^* \in \mathcal{T} \land \hat{t} = \bot
\end{equation}
This failure mode arises when $\textit{desc}_{t^*}$ exhibits insufficient specificity, yielding $P(t^* \mid q, \mathcal{M}_{t^*}) < \tau$ where $\tau$ is the selection threshold.

\noindent\textit{Failure II} (Erroneous Selection): occurs when an incorrect tool is selected:
\begin{equation}
\mathcal{F}_{II}: \hat{t} \neq \bot \land \hat{t} \neq t^*
\end{equation}
This failure mode arises from descriptions characterized by excessive generality or semantic ambiguity, such that $\exists t_j \neq t^*: P(t_j \mid q, \mathcal{M}_j) > P(t^* \mid q, \mathcal{M}_{t^*})$.

In competitive deployment scenarios wherein functionally equivalent servers coexist (i.e., $\exists t_i, t_j: \textit{impl}_i \equiv \textit{impl}_j \land \textit{desc}_i \neq \textit{desc}_j$), description quality constitutes the primary determinant of selection.

Despite the demonstrated criticality of tool descriptions to MCP system efficacy, a systematic characterization of effective description practices remains absent from the literature.
Furthermore, no empirical investigation has examined the distribution of description quality across the MCP ecosystem at scale.
These observations necessitate a rigorous investigation into the attributes that constitute effective tool descriptions, the prevalence of quality deficiencies in practice, and the feasibility of automated assessment and enhancement techniques.

\section{Constructing the Taxonomy of MCP Description Smell Standard from the Wild}


The formulation of a robust taxonomy for MCP description smells requires a methodology that bridges established software engineering (SE) principles with the unique, emerging exigencies of LLM-based tool invocation. 
Standardized descriptions are necessitated by the operational challenges of scaling tool ecosystems. 
As established in \S\ref{sec:problem}, tool selection constitutes a ``critical bottleneck that directly impacts MCP agent efficiency and autonomy''. 
Whether utilizing Retrieval-Augmented Generation (RAG) or hierarchical indexing, the efficacy of selection architectures depends fundamentally on the semantic fidelity of tool metadata. 
Inaccurate, vague, or incomplete descriptions hinder reliable function identification.
To ensure the resulting taxonomy is both theoretically grounded and empirically relevant, we employ a multi-stage synthesis of normative standards and ``in-the-wild'' observations:
\textbf{(1) Top-down theoretical foundation}: We first conduct a systematic literature review of foundational SE quality standards and API documentation research to derive a candidate list of quality dimensions (\S\ref{sec:approach:derive:paper}).
\textbf{(2) Bottom-up empirical extraction}: To capture the specific descriptive ``smells'' unique to the MCP ecosystem, we collect a large-scale corpus of 10,831 MCP servers (\S\ref{sec:approach:collect}) and extract their metadata using AST-based pattern matching (\S\ref{sec:approach:extract}).
\textbf{(3) Synthesis and mapping}: Finally, we utilize an LLM-based incremental card-sorting process to distill these raw empirical deficiencies into recurring categories, which are then mapped back to the theoretical standards to produce the final taxonomy (\S\ref{sec:approach:derive:server}).

\subsection{Deriving Potential Description Standards from the Literature Reviews}\label{sec:approach:derive:paper}

\begin{table}[]
{\footnotesize
\begin{tabular}{|p{2.3cm}|p{8.5cm}|p{1.5cm}|}
\hline
Criteria & Description & References \\
\hline\hline
\rowcolor{gray!10} Accuracy & The description needs to be consistent with the implementation of the MCP server tool & \cite{iso912612001} \cite{aghajani2019software} \cite{dautovic2011automatic}  \\
Functional & The description should clearly state the functional requirements and what it is used for it should also clearly state when the agent should be called to meet the purpose. & \cite{iso912612001} \cite{iso250102023} \cite{llamaindex} \cite{langchain} \\
\rowcolor{gray!10} Completeness & Covers all the information required by the agent to complete the task (such as parameter description, return value description, error handling, side effect, etc.) & \cite{iso265142022} \cite{iso250002005} \cite{aghajani2019software}  \cite{uddin2015api}\cite{dautovic2011automatic} \cite{aghajani2020software} \\
Concision & Avoid being too bloated and increase agent token consumption & \cite{khan2021automatic} \cite{uddin2015api} \\
\rowcolor{gray!10} Clarity & Avoid vague expressions (such as "maybe" and "probably") and ambiguity. & \cite{aghajani2019software} \\
Format Standardization & The format needs to meet the standard / whether it meets its own standards and is consistent with the server's tools desc standards & \cite{iso912612001} \cite{iso265142022}\cite{meng2020optimizing} \\
\rowcolor{gray!10} Terminology Consistency & Unify global terms within description and avoid polysemy & \cite{iso265142022} \cite{dautovic2011automatic} \\
Readability & The description should be easy to understand and learn, and should not contain overly obscure or technical words that the LLM cannot understand. & \cite{iso912612001}  \cite{ieee8301998}  \\
\hline\hline
\end{tabular}
\caption{
Quality standards suitable for MCP servers.
}
\label{tab:survey_standards}}
\end{table}


To establish a rigorous foundation for evaluating MCP server descriptions, we follow a systematic process to derive quality standards, drawing from established software engineering principles, API documentation research, and emerging requirements for LLM-based agentic systems.
We conduct a multi-stage literature review to identify attributes that define high-quality descriptive metadata. 
This process is structured into three tiers of analysis to ensure that our standards are both theoretically grounded and practically relevant to the MCP ecosystem.

\sssec{Foundational software quality standards}. The first tier focuses on established international standards for software and documentation quality. 
We adopt the core concepts of functional suitability and reliability from ISO/IEC 25010 \cite{iso250102023} and ISO/IEC 9126 \cite{iso912612001}. 
These standards emphasize that any software interface must accurately represent its capabilities and behave predictably. 
In the context of MCP, this translates to the requirement that a tool's description must faithfully reflect its underlying implementation (\textit{accuracy}) and clearly define its intended use cases (\textit{functional}). 
Additionally, we refer to IEEE Std 830 \cite{ieee8301998} for software requirements specifications, which highlights the necessity of \textit{consistency} in technical communication.

\sssec{API documentation and documentation smells}.
The second tier examines empirical research on API usability and documentation defects. 
We heavily reference the taxonomy of documentation issues by Aghajani et al. \cite{aghajani2019software, aghajani2020software}, which identifies 162 types of problems, including ambiguity and incompleteness. 
Zhong and Su \cite{zhong2013detecting} and Uddin and Robillard \cite{uddin2015api} demonstrate that inconsistencies between documentation and code lead to significant developer friction. 
Furthermore, we incorporate the concept of ``documentation smells'' from Khan et al. \cite{khan2021automatic}, specifically focusing on smells like \textit{incomplete explanations} and \textit{bloated descriptions}. 
These works justify our inclusion of completeness, clarity, and concision as essential metrics for MCP descriptions, as they directly impact how effectively a ``user'' (whether human or LLM) can interpret an interface.

\sssec{Machine-centric and agentic requirements}.
The final tier addresses the unique demands of MCP, where the primary consumers are LLMs. 
Unlike traditional documentation intended for humans, MCP descriptions must be optimized for machine reasoning and context window constraints. 
We analyze industry best practices from OpenAI's Function Calling \cite{openai2023function} and Microsoft's Semantic Kernel \cite{microsoft2024semantic}, which suggest that concise, highly structured descriptions reduce hallucination and improve tool selection accuracy. 
Research by Yuan et al. \cite{yuan2025easytool} on \textit{EasyTool} emphasizes that ``concise tool instructions'' are critical for agent performance. This tier leads to the derivation of format standardization (to ensure structural compatibility) and terminology consistency (to avoid semantic confusion during LLM embedding or reasoning).

\sssec{Summary of derived standards}.
By synthesizing these three tiers, we define seven primary quality standards suitable for MCP servers, as summarized in Table \ref{tab:survey_standards}. 
These standards serve as the theoretical lens for our subsequent analysis. 
While these criteria provide a normative framework, the loosely-constrained nature of current MCP implementations often leads to practical ``smells''.
In \S\ref{sec:rq1}, we bridge these theoretical standards with empirical observations from our dataset to develop a concrete taxonomy of MCP description smells.

\subsection{Collecting MCP Servers}\label{sec:approach:collect}

\begin{table}
\small
\caption{
Tool register methods from collected servers.
}
\begin{tabular}{l|l|r|l|l|r}
\hline
Register Method & Language & \# & Register Method & Language & \# \\
\hline\hline
\rowcolor{gray!10} @mcp.tool() & Python & 1,648 &  server.RegisterTool & JS/TS & 732\\
@mcp.list\_tools() & Python & 361 & requestHandler(ToolListHandler)
& JS/TS & 547\\
\rowcolor{gray!10} mcp.add\_tool() & Python & 892 & syncServer.addTool() & Java & 172\\
@mcp.tool()(func\_name) & Python & 92 & McpServerToolType & C\# & 54\\
\rowcolor{gray!10} FastAPI & Python & 228 &  ServerCapabilities(tools=...) & Kotlin & 27\\
server.tool & JS/TS & 3,681  & & &\\
\hline
\end{tabular}
\label{tab:register}
\end{table}

To establish a strong foundation for deriving our MCP description standards, we conducted an extensive collection of existing MCP server data from various public sources. As of March 27, we gathered data on over 10,000 MCP servers, primarily from five dominant collection points, as detailed in Fig. \ref{fig:workflow} Stage 1.

\begin{figure}
    \centering
    \begin{minipage}[]{0.49\textwidth}
        \centering
        \includegraphics[width=\textwidth]{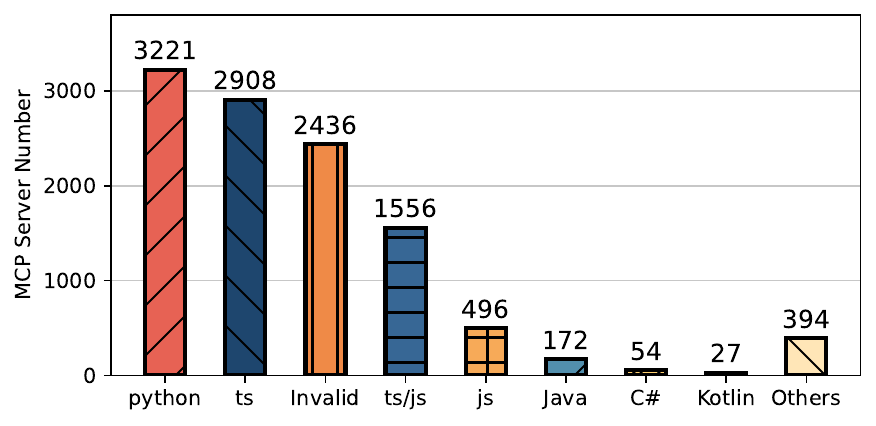}
        \caption{Distribution of MCP Server counts across different programming languages.
        }
        \label{fig:pl_count}
    \end{minipage}
    \hfill
    \begin{minipage}[]{0.49\textwidth}
        \centering
        \includegraphics[width=0.8\textwidth]{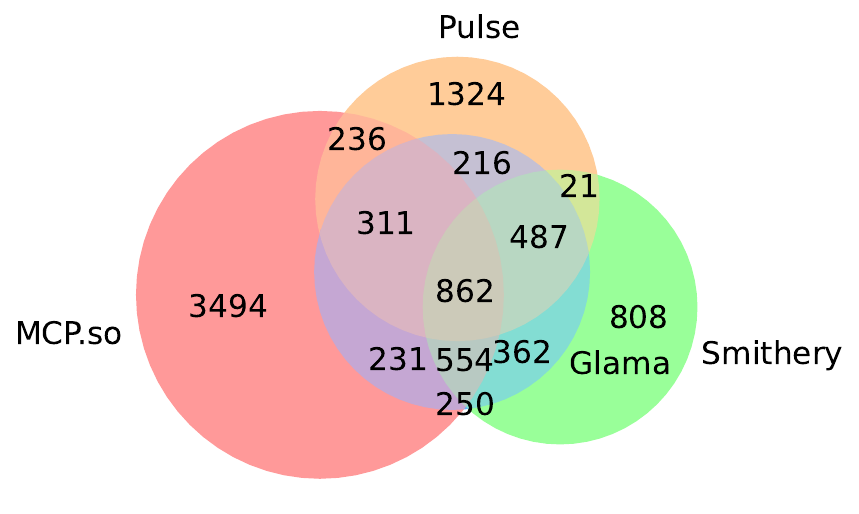}
        \caption{
        Distribution of platforms from which the servers were collected.
}
    \label{fig:platform}
    \end{minipage}
\end{figure}

To understand the overlap and unique contributions of the commercial MCP platforms (MCP.so~\cite{mcp_so}, Pulse~\cite{pulsemcp}, Glama~\cite{glama}, and Smithery~\cite{smithery}), we performed an intersection analysis. 
Fig. \ref{fig:platform} illustrates the distribution of unique and shared servers across MCP.so, Pulse, and Glama, demonstrating that a significant number of servers are unique to a single source (e.g., 3,494 servers unique to MCP.so). The intersection of all three collections yields 862 servers.

We also analyzed the programming-language distribution of the collected MCP servers to characterize the technical landscape. 
A substantial majority of the servers are written in Python (3,221 servers), mixing TypeScript/JavaScript (1,556 servers), TypeScript only (2,908 servers), or were classified as ``Invalid'' (2,436 servers). 
Other languages like JavaScript (JS), Java, C\#, and Kotlin comprise a smaller proportion of the ecosystem.

\subsection{Extracting MCP Metadata}\label{sec:approach:extract}

To enable the following analysis of the MCP servers we collected, an initial step to systematically extract their metadata (e.g., tool name, parameter, description, etc.) from the code. 

\sssec{Summarizing the register patterns of MCP servers.}
To find the metadata of registered MCP servers in their code repository, it is essential to summarize the registration patterns of MCP servers.
Firstly, since the MCP server standards allow developers to use diverse programming languages to write the code, there exist many different registration patterns from different language standards.
Furthermore, the open-source nature of the MCP source code enables the developers to specify tools using internal functions.
Therefore, we adopted an iterative checking approach, using the newly found pattern each time to try to find a registered server tool, until no new registration pattern could be found.
The registration methods we identified, along with the number of servers found using each, are summarized in the Table \ref{tab:register}.
To be noted, there exists less than 1678 servers that are registered using self-defined FastAPI standards (since MCP standards are defined based on the FastAPI standards), we do not record these servers since each of them has a diverse specification process.

\sssec{Extracting MCP server tool metadata.}
Then, based on the registration methods found, we extract the metadata from the code for next-step analysis.
The metadata extraction followed a two-stage approach: code compilation and pattern matching.
First, we transformed the source code of the collected MCP servers into an Abstract Syntax Tree (AST) structure. 
Second, we applied a pattern-matching process based on found registration patterns against the generated ASTs to identify and register every MCP server tool. 
The output of this step is the collection of a large number of description-code pairs. Each pair links a tool's natural language description (the metadata crucial for the LLM's selection process) with its corresponding function or code implementation.

\subsection{Deriving Description Standards from the Analysis of Servers}\label{sec:approach:derive:server}

\begin{algorithm}[]
\caption{LLM-Based Incremental Card Sorting for MCP Description Problems}
\footnotesize
\label{alg:card-sorting}
\begin{algorithmic}[1]
\REQUIRE Set of code--description pairs $\mathcal{P} = \{p_1, p_2, \dots, p_N\}$, sim. threshold $\tau_{\text{new}}$
\ENSURE Final taxonomy of description problem categories $\mathcal{C}$
\STATE Initialize category set $\mathcal{C} \leftarrow \emptyset$
\FOR{each pair $p_i = (d_i, c_i)$ in $\mathcal{P}$}
    \STATE \textbf{Problem Extraction:} Use the LLM to identify problems $P(p_i) = \{q_1, q_2, \dots, q_m\}$ between $d_i$ and $c_i$
    \FOR{each problem $q_k$ in $P(p_i)$}
        \STATE Compute similarity scores $s(q_k, C_j)$ for all $C_j \in \mathcal{C}$
        \IF{$\max_j s(q_k, C_j) > \tau_{\text{new}}$}
            \STATE Assign $q_k$ to best-matching category $C_j$ and increment $C_j.\textit{support}$
        \ELSE
            \STATE Create new category $C_{\text{new}}$ with (name, definition, signature, examples)
            \STATE Append $C_{\text{new}}$ to $\mathcal{C}$
        \ENDIF
    \ENDFOR
\ENDFOR
\STATE Return $\mathcal{C}$
\end{algorithmic}
\end{algorithm}

After extracting all code-description pairs from the MCP repositories, we employ a \textit{LLM-assisted card sorting process} to uncover the most common types of description problems.
The purpose of this step is to transform thousands of heterogeneous tool descriptions into a structured taxonomy of recurring issues, which later inform the MCP Description Standards.

Following the steps in Algorithm \ref{alg:card-sorting}, given a set of code-description pairs $\mathcal{P} = \{p_1, p_2, \dots, p_N\}$, each pair $p_i = (d_i, c_i)$ contains the textual description $d_i$ and its corresponding implementation $c_i$. 
We aim to iteratively analyze these pairs to identify problems such as inaccurate summaries, incomplete parameter coverage, or vague behavioral explanations. 
The algorithm maintains an evolving set of problem categories $\mathcal{C}$ and continuously refines it as more examples are processed.

The procedure begins with an empty category set $\mathcal{C} = \emptyset$. 
For each pair, the LLM first compares $d_i$ and $c_i$ to detect concrete inconsistencies or omissions, producing a list of candidate problems $P(p_i) = \{q_1, q_2, \dots, q_m\}$. 
Each candidate problem is then matched against the existing category set $\mathcal{C}$ using semantic similarity. 
If a match with confidence above a predefined threshold $\tau_{\text{new}}$ exists, the problem is assigned to that category. 
Otherwise, the model creates a new category, with a brief definition, name, and supporting evidence from the pair. 
Each category $C_j$ maintains attributes:
\begin{itemize}[leftmargin=*]
    \item \textbf{Name:} concise label describing the issue (e.g., ``Description--Code Inconsistency'');
    \item \textbf{Definition:} one-sentence summary of the underlying problem;
    \item \textbf{Signature:} textual or structural cues indicating the problem type;
    \item \textbf{Examples:} representative code--description pairs;
    \item \textbf{Support:} frequency count and confidence estimate.
\end{itemize}
The detailed LLM-driven procedure is formalized in Algorithm~\ref{alg:card-sorting}, using the model gpt-4o-mini. 
At last, we manually map the result category to potential description standards derived in \S\ref{sec:approach:derive:paper}.
Clarity, format standardization, terminology consistency, and readability are not mapped in the final results.
Specifically, since the MCP server tool mainly focuses on describing tool functions, there is no need to state any vague expressions (such as ``maybe'' and ``probably'').
Moreover, since there has not been a standard yet, format standardization has not been identified as a smell yet.
Next, most descriptions are short, which makes it easy to keep terminology consistent.
Lastly, readability is not mapped since it is a standard for humans more than LLMs.

\section{Empirical Study on Description Smells in the Wild}

This section presents the results of a study of the MCP description smells in the wild stated in \S\ref{sec: approach}.
\S\ref{sec:rq1} presents the taxonomy of MCP description smells \& standards, while \S\ref{sec:rq2} analyzes the prevalence of smells.


\begin{table*}[]
\centering
\footnotesize
\caption{In-wild standard for description of MCP servers.}
\label{tab:standard}
\renewcommand{\arraystretch}{1.15}
\scalebox{0.85}{
\begin{tabular}{l|l|r|l|l|r}
\hline
\rowcolor[HTML]{EFEFEF}
\textbf{Sub Category} & \textbf{Problem} & \textbf{\#}
& \textbf{Sub Category} & \textbf{Problem} & \textbf{\#} \\
\hline\hline

\multicolumn{3}{c|}{\textbf{Accuracy}}
& \multicolumn{3}{c}{\textbf{Information completeness}} \\
\hline

\multirow{4}{*}{Functional logic match}
& \cellcolor[HTML]{EFEFEF}Inconsistent behavior & \cellcolor[HTML]{EFEFEF}72
& \multirow{3}{*}{Complete parameters}
& \cellcolor[HTML]{EFEFEF}No desc.\ of param. & \cellcolor[HTML]{EFEFEF}1,285 \\

& Unclear boundaries & 1,137
& & No desc.\ of param.\ type & 14 \\

& \cellcolor[HTML]{EFEFEF}Non-existed behavior & \cellcolor[HTML]{EFEFEF}572
& & \cellcolor[HTML]{EFEFEF}No desc.\ of return values & \cellcolor[HTML]{EFEFEF}3,093 \\
\cline{4-6}

& Undeclared behavior & 931
& \multirow{3}{*}{Complete inner info.}
& \multirow{3}{*}{Missing side effects} & \multirow{3}{*}{773} \\
\cline{1-3}

\multirow{2}{*}{Parameter match}
& \cellcolor[HTML]{EFEFEF}Wrong param.\ type & \cellcolor[HTML]{EFEFEF}203
& & & \\

& Wrong param.\ meaning & 3,449
& & & \\
\hline

\multicolumn{3}{c|}{\textbf{Functionality}}
& \multicolumn{3}{c}{\textbf{Conciseness}} \\
\hline

\multirow{2}{*}{Clear functional description}
& \cellcolor[HTML]{EFEFEF}Repeated tool names & \cellcolor[HTML]{EFEFEF}7,894
& \multirow{2}{*}{No redundant information}
& \cellcolor[HTML]{EFEFEF}Repeated functional description & \cellcolor[HTML]{EFEFEF}154 \\

& Confusing functional desc. & 3,572
& & Clutter with irrelevant details & 2,904 \\
\hline

\multirow{2}{*}{Trigger conditions}
& \cellcolor[HTML]{EFEFEF}No trigger conditions & \cellcolor[HTML]{EFEFEF}2,972
& \multirow{2}{*}{No complex information}
& \cellcolor[HTML]{EFEFEF}Overuse of technical terms & \cellcolor[HTML]{EFEFEF}467 \\

& Confusing trigger conditions & 1,583
& & Useless qualifiers & 324 \\
\hline

\end{tabular}}
\end{table*}

\subsection{RQ1: The Taxonomy of MCP Description Smells \& Standards}\label{sec:rq1}



Based on the results we obtained \S\ref{sec:approach:derive:server}, we analyzed the corpus of 10,831 MCP server instances to identify recurring patterns of defective documentation. 
We term these patterns MCP description smells, characteristics in the tool metadata that hinder an LLM's ability to select or call tools correctly. 
By synthesizing these observed deficiencies, we derived four critical dimensions for high-quality MCP descriptions: accuracy, functionality, information completeness, and conciseness. 
Table \ref{tab:standard} presents the mapping between the identified smells and the derived standards.

\sssec{Accuracy}. 
This dimension ensures the semantic alignment between the natural language description and the underlying code implementation. 
Our analysis uncovered critical smells such as inconsistent behavior and unclear boundaries. 
For instance, if a description implies capabilities that contradict the actual logic, the agent may attempt invalid operations. 
To mitigate these risks, the standard requires a strict functional logic match and parameter match. 
Specifically, descriptions must accurately reflect the tool's behavioral limits to prevent the LLM from executing non-existent behaviors or misinterpreting argument constraints.

\sssec{Functionality}. 
Functionality addresses the distinctiveness of a tool within a shared namespace. 
A prevalent smell identified was repeated tool names and confusing functional descriptions, where multiple tools shared identical or semantically overlapping identifiers. 
This ambiguity causes the LLM to hallucinate or select tools arbitrarily. 
Consequently, the derived standard mandates explicit trigger conditions. 
A robust description must not only state what the tool does but also explicitly define the context in which it should be prioritized over others to ensure clear functional boundaries.

\sssec{Information completeness}. 
This dimension ensures the agent possesses all necessary metadata to construct a valid API call and interpret the result without ``guessing''.
We identified frequent omissions, specifically, no description of return values and missing side effects. 
The absence of output schema definitions forces the LLM to operate in a zero-shot manner regarding data processing, significantly increasing the likelihood of downstream failure. 
Therefore, the standard requires the comprehensive provision of complete parameters and complete inner information (e.g., error handling, side effects) to support autonomous reasoning loops.

\sssec{Conciseness}. 
Finally, this dimension optimizes the signal-to-noise ratio of the prompt context. Our analysis revealed smells related to clutter with irrelevant details and repeated functional descriptions. 
Excessive verbosity or the overuse of technical terms inflates token consumption and dilutes the model's attention mechanism. 
The standard advocates for no redundant information and no complex information, ensuring that the metadata provides sufficient context for task completion without inducing prompt bloat.

\finding{The proposed MCP description standard comprises four dimensions (accuracy, functionality, completeness, and conciseness) derived from 18 specific problem categories identified through large-scale code-description analysis.}


\subsection{RQ2: The State of Practice: Prevalence of Smells}\label{sec:rq2}

To assess the state of practice in the MCP ecosystem, we applied the standards-based assessment criteria derived in RQ1 to the collected corpus of 10,831 MCP servers. 
The measurement reveals a concerning prevalence of description smells, indicating that a significant portion of the ecosystem currently fails to meet the requirements for reliable agentic interoperability.

\sssec{Prevalence of functionality ambiguity}. 
The analysis identifies Functionality as the most compromised dimension. 
As shown in Table \ref{tab:standard}, repeated tool names constitute the single most dominant failure, appearing in 7,894 cases (approx. 73\% of the corpus). 
This high frequency suggests that many developers name tools based on generic internal utility functions (e.g., read\_file, get\_data) rather than unique, semantically distinct identifiers required for a shared agent namespace. 
Furthermore, 3,572 instances exhibit confusing functional descriptions, and 1,137 cases suffer from unclear boundaries. 
In a multi-server environment, this semantic ambiguity effectively prevents the LLM from distinguishing between competing tools, leading to arbitrary selection behaviors.

\sssec{Critical gaps in accuracy and completeness}. 
While functionality issues are the most numerous, deficiencies in accuracy and information completeness pose the most severe risks to execution reliability. 
We observed 3,449 instances of wrong parameter meaning, where the natural language description fails to align with the underlying schema or code logic. 
This discrepancy indicates a ``Code-First, Description-Last'' development pattern, where documentation is treated as an afterthought rather than a functional interface contract.

Moreover, the ecosystem suffers from a lack of output specifications. We identified 3,093 cases with no description of return values. 
This omission is particularly detrimental for autonomous agents; without knowledge of the output structure (e.g., JSON schema or text format), the LLM is forced to operate in a zero-shot manner when processing tool results. This significantly increases the probability of hallucination during subsequent reasoning steps.

\sssec{Inefficiency via verbosity}. 
Finally, regarding conciseness, 2,904 tools were flagged for clutter with irrelevant details. While less critical for correctness, this widespread ``prompt bloat'' introduces unnecessary noise into the model's context window, increasing inference costs and potentially degrading attention on critical instructions.

\finding{Current MCP repositories suffer from severe documentation smells, primarily dominated by repeated tool names (7,894), wrong parameter meanings (3,449), and missing return value descriptions (3,093), all of which degrade the performance of LLM-based agents.}
\section{Proposed Standard Effectiveness}
To rigorously evaluate the effectiveness of our proposed MCP server description standard, we adopt a systematic controlled variable methodology that enables precise measurement of individual component contributions while accounting for real-world competitive dynamics.

As established in the standard we proposed in Section~\ref{sec:rq2} (Table~\ref{tab:standard}), a ``good'' MCP server description covers four dimensions: \textit{Accuracy}, \textit{Functionality}, \textit{Information Completeness}, and \textit{Conciseness}. To measure the effectiveness of these dimensions, both individually and in real-world competitive scenarios, we pose two research questions:

\begin{itemize}[leftmargin=*]
    \item RQ3: What is the individual contribution of each description dimension (Accuracy, Functionality, Information Completeness, Conciseness) to the overall tool selection process? 
    \item RQ4: Does adhering to our proposed description standard provide a competitive advantage in real-world scenarios where multiple functionally equivalent MCP servers coexist? 
\end{itemize}

\paragraph{Experimental Setup.} We enforce a stateless evaluation environment by initializing a new session for each query. This isolation prevents the model from relying on interaction history or favoring previously selected tools. Within each session, we assign a randomly generated UUID to each tool. This configuration serves two purposes: it enables the simultaneous deployment of multiple MCP servers without namespace conflicts, and it ensures that tool selection is based solely on description quality, independent of naming semantics or prior context.
We utilize gpt-4o-mini as the base model for the experiment.

\subsection{RQ3: Individual Component Contribution Analysis}
\label{ref:rq3}

Building upon the four-dimensional MCP description standard derived in RQ1 (Table~\ref{tab:standard}), namely \textit{Accuracy}, \textit{Functionality}, \textit{Information Completeness}, and \textit{Conciseness}, and the prevalent description smells identified in RQ2, we now investigate how each dimension individually contributes to LLM tool selection behavior. This analysis directly addresses whether the quality issues we documented in practice (e.g., wrong parameter meanings, repeated tool names, missing return value descriptions) translate into measurable selection failures.

To identify which description standard dimension most significantly influences tool selection probability, we employ a dual-mutation experimental design that systematically varies both tool descriptions and task queries. The mutation algorithm is shown in Algorithm~\ref{alg:mcp_mutation}. Specifically, it operates in two stages to generate mutated descriptions and queries while maintaining implementation consistency.

In \textbf{Stage 1}, we generate $n$ server variants, each emphasizing a different description dimension from our proposed standard (Table~\ref{tab:standard}). The \textsc{GenerateDescription} function prompts an LLM to create a description for server $S$ based on its code implementation and specific dimension requirements $d_i$. Each dimension $d_i$ corresponds to one of our four standard categories:
\begin{itemize}[leftmargin=*]
    \item \textit{Accuracy}: ensuring functional logic match and parameter match, addressing issues such as inconsistent behavior, unclear boundaries, and wrong parameter meaning.
    \item \textit{Functionality}: providing clear functional descriptions and explicit trigger conditions, mitigating problems like repeated tool names and confusing functional descriptions.
    \item \textit{Information Completeness}: covering all parameters, return values, and side effects, resolving issues such as missing parameter descriptions and undocumented return values.
    \item \textit{Conciseness}: eliminating redundant information and complex terminology, avoiding clutter with irrelevant details and overuse of technical terms.
\end{itemize}
The resulting description $\mathrm{Desc}_i$ is then paired with the original server code to create a variant $S_i$, ensuring that all servers are functionally identical but differ only in their documentation quality along one dimension.

In \textbf{Stage 2}, we generate a diverse set of queries that test the robustness of each description across varying \textit{query complexity levels}, which quantify how much detail and specificity a user provides when requesting a tool. The \textsc{GenerateQueries} function produces $m=100$ query mutations based on the server's functionality, with 10 queries generated at each of the 10 complexity levels (Table~\ref{tab:query_complexity}). Level 1 represents minimal queries (brief, underspecified) while Level 10 represents complete queries (detailed, fully specified). This gradient captures realistic user behavior, where queries may vary in clarity and completeness.


We apply this mutation to 10 distinct types of MCP servers, spanning diverse functionalities including data processing, file operations, API integrations, and computational tasks. For each server type, we generate description variants across all dimensions and corresponding query sets. 

\begin{table}[]
\centering
\caption{Query complexity levels (1--10) from underspecified to fully specified.}
\label{tab:query_complexity}
\footnotesize
\begin{tabular}{@{}cp{0.38\textwidth}|cp{0.38\textwidth}@{}}
\toprule
\textbf{Level} & \textbf{Example Query} & \textbf{Level} & \textbf{Example Query} \\
\midrule
1 & {Read file.} & 2 & {Read config.json.} \\
\midrule
3 & {Read the config file.} & 4 & {Read the configuration file in the project.} \\
\midrule
5 & {Can you read the config file? I need to check settings.} & 6 & {Read the configuration file located in the project root directory.} \\
\midrule
7 & {I need to read the config file to check the database settings.} & 8 & {Read the config file at /var/www/myapp/config.json to check parameters.} \\
\midrule
9 & {I'm debugging an issue. Read /var/www/myapp/config/database.json for connection settings.} & 10 & {I'm troubleshooting a production issue. Read the config file at /var/www/myapp/config/database.json to check connection parameters and timeout values.} \\
\bottomrule
\end{tabular}
\end{table}

\begin{table}[]
\centering
\caption{Impact of each description dimension on tool selection probability across query complexity levels.}
\label{tab:dimension_selection_stats}
\footnotesize
\begin{tabular}{lcccc|lcccc}
\toprule
\textbf{Dimension} & \textbf{With} & \textbf{W/O} & \textbf{$\Delta$} & \textbf{$p$} & \textbf{Dimension} & \textbf{With} & \textbf{W/O} & \textbf{$\Delta$} & \textbf{$p$} \\
\midrule
\multicolumn{5}{l|}{\textit{Overall (n=1,000)}} & \multicolumn{5}{l}{\textit{Minimal Queries (Level 1-3, n=300)}} \\
Functionality & 17.8\% & 6.2\% & +11.6\% & *** & Functionality & 19.2\% & 5.0\% & +14.2\% & *** \\
Accuracy & 15.9\% & 7.1\% & +8.8\% & *** & Accuracy & 14.1\% & 7.8\% & +6.3\% & ** \\
Info. Completeness & 14.3\% & 8.4\% & +5.9\% & ** & Info. Completeness & 12.5\% & 8.9\% & +3.6\% & * \\
Conciseness & 10.6\% & 9.1\% & +1.5\% & * & Conciseness & 9.8\% & 9.0\% & +0.8\% & -- \\
\midrule
\multicolumn{5}{l|}{\textit{Moderate Queries (Level 4-7, n=400)}} & \multicolumn{5}{l}{\textit{Complete Queries (Level 8-10, n=300)}} \\
Functionality & 17.5\% & 6.4\% & +11.1\% & *** & Functionality & 16.7\% & 7.2\% & +9.5\% & *** \\
Accuracy & 17.2\% & 7.1\% & +10.1\% & *** & Accuracy & 16.4\% & 6.4\% & +10.0\% & *** \\
Info. Completeness & 14.0\% & 8.2\% & +5.8\% & ** & Info. Completeness & 16.8\% & 8.5\% & +8.3\% & *** \\
Conciseness & 10.9\% & 9.2\% & +1.7\% & * & Conciseness & 11.1\% & 9.1\% & +2.0\% & * \\
\midrule
\multicolumn{10}{l}{\textit{Baseline: 10.0\% (random selection among 10 servers). *$p<$0.05, **$p<$0.01, ***$p<$0.001, -- not significant.}} \\
\bottomrule
\end{tabular}
\end{table}

\begin{algorithm}
\caption{Description and Query Mutation Generation}
\footnotesize
\label{alg:mcp_mutation}
\begin{algorithmic}[1]
\STATE \textbf{Input:} 
\STATE \hspace{1em} MCP server $S$ with fixed code implementation
\STATE \hspace{1em} Description dimensions $D = \{d_1, d_2, \ldots, d_n\}$
\STATE \hspace{1em} Number of query mutations $m$
\STATE \textbf{Output:} 
\STATE \hspace{1em} Query set $Q = \{q_1, q_2, \ldots, q_m\}$
\STATE \hspace{1em} MCP server set $\mathcal{S} = \{S_1, S_2, \ldots, S_n\}$
\STATE
\STATE \textbf{Stage 1: Description Generation}
\STATE Initialize $\mathcal{S} \leftarrow \emptyset$
\FOR{ $d_i \in D$}
    \STATE $\mathrm{Desc}_i \leftarrow \textsc{GenerateDescription}(S, d_i)$
    \STATE $S_i \leftarrow \textsc{CreateServer}(S.\mathrm{code}, \mathrm{Desc}_i)$
    \STATE $\mathcal{S} \leftarrow \mathcal{S} \cup \{S_i\}$
\ENDFOR
\STATE
\STATE \textbf{Stage 2: Query Generation}
\STATE $Q \leftarrow \textsc{GenerateQueries}(S, m)$
\STATE
\STATE \textbf{Return} $Q$ and \ensuremath{\mathcal{S}}
\end{algorithmic}
\end{algorithm}

Table~\ref{tab:dimension_selection_stats} presents the results of our ablation study across 10 MCP servers with 100 queries each (1,000 total queries). For each dimension, we compare tool selection rates when the dimension is well-implemented (With) versus poorly-implemented (W/O), while keeping other dimensions constant. We report statistical significance using chi-squared tests.

\paragraph{Overall Dimension Impact.}
Among all four dimensions, \textit{Functionality} exhibits the strongest impact on tool selection, with a $\Delta$ of +11.6\% (17.8\% vs 6.2\%, $p < 0.001$). 

\textit{Accuracy} ranks second with a $\Delta$ of +8.8\% ($p < 0.001$), highlighting the importance of precise parameter documentation. This aligns with our identification of wrong parameter meaning (3,449 cases) as a critical issue in RQ2.

\textit{Information Completeness} shows moderate but significant impact (+5.9\%, $p < 0.01$), confirming that missing return value descriptions (3,093 cases) and parameter descriptions (1,285 cases) do affect selection reliability.

\textit{Conciseness} (+1.5\%, $p < 0.05$) demonstrates smaller but still statistically significant effects, suggesting that while clutter with irrelevant details (2,904 cases) is prevalent, its impact on selection is less severe than other dimensions.

\paragraph{Impact by Query Complexity.}
We further analyze how query complexity affects dimension importance by grouping the 10 complexity levels into three categories: Level 1-3 (minimal), Level 4-7 (moderate), and Level 8-10 (complete).

For \textbf{minimal queries (Level 1-3)}, \textit{Functionality} shows the largest effect ($\Delta$ = +14.2\%, $p < 0.001$), as underspecified requests offer few clues beyond the tool's core purpose. \textit{Accuracy} follows with +6.3\% ($p < 0.01$), while \textit{Information Completeness} shows reduced impact (+3.6\%, $p < 0.05$). Notably, \textit{Conciseness} shows no significant effect at this level ($p > 0.05$), indicating that brevity matters less when queries are already minimal.

For \textbf{moderate queries (Level 4-7)}, both \textit{Functionality} ($\Delta$ = +11.1\%, $p < 0.001$) and \textit{Accuracy} ($\Delta$ = +10.1\%, $p < 0.001$) become equally important. At this complexity level, users provide enough context that parameter details help distinguish between similar tools. All four dimensions show significant effects.

For \textbf{complete queries (Level 8-10)}, \textit{Information Completeness} gains the most influence ($\Delta$ = +8.3\%, $p < 0.001$), matching the richer context provided by users who expect comprehensive documentation. \textit{Accuracy} remains strong (+10.0\%, $p < 0.001$), while \textit{Functionality}'s impact slightly decreases (+9.5\%) as users' detailed queries already convey intent.

\finding{Our analysis across 10 server types and four description dimensions reveals that \textit{Functionality} has the strongest overall impact on tool selection (+11.6\%, $p < 0.001$), followed by \textit{Accuracy} (+8.8\%), \textit{Information Completeness} (+5.9\%), and \textit{Conciseness} (+1.5\%). The relative importance shifts with query complexity: \textit{Functionality} dominates for minimal queries (Level 1-3), while \textit{Information Completeness} becomes critical for complete queries (Level 8-10). These results further validate that the description smells identified in RQ2 cause measurable selection failures.}

\subsection{RQ4: Real-World Validation of Standard-Compliant Descriptions}
\label{ref:rq4}

While RQ3 demonstrates the individual contribution of each dimension through controlled mutations on synthetic description variants, in this research question, we further validate whether our proposed standard provides practical benefits in real-world competitive scenarios. In practice, the MCP ecosystem contains multiple servers offering similar functionality. For example, numerous file system servers, database connectors, and browser automation tools coexist and compete for selection by LLM-based agents. When a user issues a query, the LLM must choose among these functionally equivalent alternatives based primarily on their descriptions. This research question investigates whether a server that adheres to our four-dimensional standard (Table~\ref{tab:standard}) gains a measurable competitive advantage over servers with their original, unoptimized descriptions.

To conduct this evaluation, we collected groups of functionally equivalent MCP servers from existing platforms, selecting 10 groups across diverse domains: FileSystem, Terminal Controller, Browser Automation, OS Automation, Finance, Communication, Calendar Management, Cloud Storage, Location Services, and Databases. Each group contains five real servers deployed in the wild with their original implementations and descriptions intact. Within each group, we optimized \textit{only one} server's description to fully comply with our proposed standard by correcting \textit{Accuracy} issues (inconsistent behavior descriptions and wrong parameter meanings), adding clear \textit{Functionality} descriptions and explicit trigger conditions, ensuring \textit{Information Completeness} by documenting all parameters, return values, and side effects, and improving \textit{Conciseness} by removing redundant information and simplifying technical terminology. The other four servers in each group retained their original descriptions completely unchanged. This setup mirrors the real-world scenario where one server developer invests effort in documentation quality while competitors do not.

We generated $m=100$ queries per group using the same methodology as RQ3 (Algorithm~\ref{alg:mcp_mutation}), with 10 queries at each complexity level (Level 1-10), resulting in a total of 1,000 queries across all groups. For each query, all five servers in the group were presented simultaneously to the LLM, which selected one server to handle the task. We measured the selection probability for each server and computed the competitive advantage as the relative increase over the uniform baseline ($P_{\text{baseline}} = 1/5 = 20\%$).

Across all 10 server groups, the standard-compliant server achieved an average selection probability of 72\%, compared to the 20\% baseline expected under uniform random selection. This represents a 260\% relative increase in selection probability, demonstrating that description quality has a substantial impact on competitive outcomes. The advantage was consistent across all tested domains, with selection probabilities ranging from 65\% (Location Services) to 81\% (FileSystem). 

We further analyzed the results by query complexity to understand how user behavior affects competitive dynamics. For minimal queries (Level 1-3), the optimized server achieved 78\% selection probability, as underspecified queries heavily rely on clear functional descriptions to disambiguate among similar tools. For moderate queries (Level 4-7), the selection probability was 71\%, with accurate parameter documentation becoming increasingly important. For complete queries (Level 8-10), the advantage slightly decreased to 68\%, but remained substantial, indicating that comprehensive documentation benefits users regardless of query specificity. This pattern aligns with our RQ3 findings, where \textit{Functionality} dominated for minimal queries while \textit{Information Completeness} gained importance for complete queries.

\finding{In real-world competitive scenarios with five functionally equivalent servers, the standard-compliant server achieves a 72\% selection probability, representing a 260\% increase over the 20\% baseline ($p < 0.01$). This advantage is consistent across all 10 tested domains (ranging from 65\% to 81\%) and all query complexity levels (68\%-78\%). These results demonstrate that adhering to our proposed description standard provides a significant competitive advantage in the MCP ecosystem, offering practical guidance for server developers seeking to maximize tool adoption.}
\section{Related Work}

\textbf{Tool Selection in LLM Agents.}
Recent works have explored the security implications of tool use and tool selection in LLM-based agents.
A line of research focuses on preventing inappropriate or unsafe tool invocations caused by adversarial or untrusted inputs, often studied in the context of prompt injection and related attacks~\cite{wu2024isolategpt, zhan2024injecagent, iqbal2024llm, fu2024imprompter, debenedetti2024agentdojo, hines2024defending, bagdasarian2024airgapagent}.
These approaches typically model violations as deviations of agent behavior from the user’s intended goals due to maliciously crafted prompts or compromised contextual inputs, and propose defenses such as execution isolation, input sanitization, or constrained tool invocation to mitigate such risks. As a result, their primary focus lies in controlling the agent’s tool selection and invocation flow to prevent unintended or unsafe executions. 
Another line of work further examines the mechanisms underlying tool selection in LLM-based agents~\cite{lumer2025tool, masterman2024landscape, jia2025autotool, yuan2025easytool, chen2024towards}. Lumer et al.~\cite{lumer2025tool} claim that APIs are typically designed for human developers to invoke explicitly, while tools are optimized for agents that autonomously decide when and how to use them based on their name, description, and parameter schema. MCP tools, in particular, standardize this interface by
defining reusable, interoperable tool specifications that can be hosted remotely or run locally. EASYTOOL~\cite{yuan2025easytool} reformulates and compresses tool documentation into concise, standardized tool instructions to better guide LLMs in selecting appropriate tools and constructing valid tool invocations.
In contrast, while these works aim to improve tool selection by refining general-purpose tool documentation or instructions, our work focuses on how MCP tool calling descriptions should be designed and standardized as a protocol-level interface, grounded in large-scale empirical analysis of existing MCP servers.


\noindent\textbf{API Documentations.}
The quality of API documentation is a critical factor in software ecosystems and has been the subject of extensive empirical research. Seminal studies \cite{uddin2015api, dautovic2011automatic}. 
established that documentation obstacles, particularly severe incompleteness and ambiguity, serve as primary barriers to effective API adoption and learning. 
Building on this foundation, previous researchers differentiated between content-related problems (e.g., incorrectness, obsoleteness) and presentation problems (e.g., bloat, fragmentation), finding that while developers prioritize accurate content, poor presentation significantly hinders productivity \cite{stocker2024pattern, serbout2021openapi}. 
This classification was further refined by Aghajani et al., who mined software repositories to construct a comprehensive taxonomy of 162 documentation issue types, spanning information content, usability, and maintenance processes \cite{aghajani2019software, aghajani2020software}. 
Concurrently, researchers have sought to operationalize these quality attributes; for instance, Zhong et al. emphasized the systematic assessment of completeness and consistency \cite{zhong2013detecting}, while Khan et al. recently demonstrated the feasibility of automatically detecting specific ``documentation smells'', such as ``lazy'' or ``tangled'' descriptions, using deep learning techniques \cite{khan2021automatic}. 
Our work is directly inspired by these established taxonomies and the methodological progression toward automated quality analysis. 
However, unlike prior studies that focus predominantly on human consumption of standard library APIs, we apply these rigorous quality perspectives to the MCP servers. 
In the MCP context, documentation serves a dual purpose: it must be intelligible to human developers while simultaneously functioning as a precise context for LLMs. 
We therefore adapt the concepts of documentation correctness and completeness to evaluate how well the MCP server specifications support reliable and hallucination-free model interactions.
\section{Discussion}

\subsection{Implication}

Our findings carry several implications for MCP server developers, platform maintainers, and the broader agent ecosystem.

\sssec{Description quality is a first-class engineering concern, not an afterthought.}
Our large-scale measurement (RQ2) reveals that the majority of MCP servers suffer from at least one category of description smell, with repeated tool names alone affecting approximately 73\% of the corpus.
Combined with the experiments in RQ3 and RQ4, which demonstrate that description quality directly determines tool selection probability, our results establish that writing a high-quality description is not a cosmetic documentation task but a functional engineering requirement.
In particular, the ``Code-First, Description-Last'' development pattern we observed, where developers prioritize implementation and treat documentation as an afterthought, leads to measurable degradation in tool discoverability and invocation accuracy.
We therefore advocate that MCP server development workflows should treat description authoring as a co-equal phase alongside implementation and testing, similar to how API contract design is treated in modern service-oriented architectures.

\sssec{From description quality to agent skill acquisition.}
Beyond the MCP ecosystem, our findings have broader implications for how LLM-based agents acquire and compose skills.
In modern agentic platforms, the concept of \textit{Agent Skills} has emerged as a first-class abstraction for extending agent capabilities.
For example, Anthropic defines Agent Skills as modular, filesystem-based capability packages that bundle instructions, metadata, and optional resources (scripts, templates, reference materials) to transform general-purpose agents into domain specialists~\cite{anthropic-skills}.
Crucially, the discovery and activation of a Skill depends on a two-level metadata mechanism: a lightweight description (name and summary) is loaded at startup for the agent to decide \textit{when} to invoke the Skill, while detailed instructions and bundled resources are loaded on-demand only after the Skill is triggered.
This progressive disclosure architecture means that the quality of the description metadata directly determines whether an agent can discover and activate the correct Skill in the first place.
Therefore, the findings on the MCP tool descriptions map directly onto this Skill discovery problem.

\subsection{Limitations}

This study concentrates on the description metadata of MCP server tools (names, textual descriptions, and input/output schemas), while MCP servers also expose \textit{resources} and \textit{prompts}, whose quality may equally affect agent behavior; we leave the systematic analysis of these additional components to future work.
Furthermore, our measurement (RQ2) assesses description quality through static analysis of code-description pairs, without executing the tools at runtime.
Consequently, certain dynamic issues, such as descriptions that are accurate at the time of writing but become stale after code updates, are not captured by our approach.
A longitudinal study tracking description drift over time would complement our cross-sectional findings.

\subsection{Threats to Validity}

\sssec{Internal validity.}
A primary threat to internal validity concerns the experiment design in RQ3 and RQ4.
In RQ3, we generate description variants by emphasizing one quality dimension at a time while keeping others constant.
However, in practice, the four dimensions are not fully independent; improving one dimension may implicitly affect another.
For example, adding missing parameter descriptions may also improve Accuracy by correcting previously ambiguous parameter semantics.
To mitigate this, we carefully reviewed a random sample of generated variants to ensure that each mutation primarily targets its intended dimension.
However, we optimized one server's description per group while leaving the other four unchanged.
The selection advantage observed could be partially attributed to the optimized description being longer or shorter than competitors, rather than qualitatively better.
We control for this by ensuring that the optimized descriptions do not significantly differ in token length from the originals.

\sssec{External validity.}
Our dataset is collected from five public MCP platforms (GitHub, MCP.so, Glama, PulseMCP, and Smithery) as of March 2025.
This snapshot may not represent the full diversity of the MCP ecosystem, particularly enterprise-internal or proprietary servers that are not publicly registered.
Furthermore, the MCP ecosystem is evolving rapidly; the distribution of description quality may shift as the protocol matures and community best practices emerge.
The programming language  in our corpus is dominated by Python and TypeScript/JavaScript, which may limit the generalizability of our AST-based extraction patterns to servers written in less common languages.

\section{Conclusion}

To summarize, in this paper, we present the first systematic study of smells in tool descriptions in the MCP server ecosystem.
By conducting a comprehensive study on 14 standards on code and document quality, and analyzing 10,831 real-world MCP servers, we identify widespread description deficiencies and distill them into a four-dimensional quality standard covering accuracy, functionality, information completeness, and conciseness.
Through controlled ablation experiments, we demonstrate that these dimensions have a direct and statistically significant impact on LLM tool selection behavior, with functionality and accuracy playing dominant roles under underspecified queries, and completeness becoming critical as query specificity increases.
Furthermore, our real-world competitive evaluation shows that MCP servers adhering to the proposed standard achieve up to a 260\% increase in selection probability when coexisting with functionally equivalent alternatives.
Together, these results establish MCP tool descriptions as a first-class engineering interface rather thanauxiliary documentation, and provide concrete, empirically grounded guidance for building reliable, interoperatble and scalable agent ecosystems.

\section*{Data Availability}
The dataset and artifact of this paper are available at \url{https://anonymous.4open.science/r/MCP-Desc-Standard/}.

\bibliographystyle{ACM-Reference-Format}
\bibliography{acmart, survey_api, tool_selection}

\newpage

\end{document}